\documentstyle[aps,graphicx,amsmath,amssymb,bm]{revtex}

\newcommand{\kf}{k_{F}}
\newcommand{\kfn}{k_{F_n}}
\newcommand{\kfp}{k_{F_p}}
\newcommand{\kfns}{k_{F_n}^2}
\newcommand{\beq}{\begin{equation}}
\newcommand{\eeq}{\end{equation}}
\newcommand{\beqy}{\begin{eqnarray}}
\newcommand{\eeqy}{\end{eqnarray}} 
\newcommand{\q}{{\bm q}}
\newcommand{\qp}{{\bm q'}}
\newcommand{\Pt}{{\bm P}}

\newcommand{\qh}{\widehat{{\bm q}}}
\newcommand{\qph}{\widehat{{\bm q}}'}
\newcommand{\Pth}{\widehat{{\bm P}}}

\newcommand{\si}{{\bm \sigma_1}}
\newcommand{\sip}{{\bm \sigma_2}}
\newcommand{\vlk}{V_{\text{low}\,k}}

\begin{document}
\draft
\title{Neutrino Bremsstrahlung in Neutron Matter from Effective Nuclear
Interactions}

\author{Achim Schwenk$^1$, Prashanth Jaikumar$^2$ and Charles Gale$^2$}
\address{$^1$Department of Physics, The Ohio State University, Columbus, 
OH 43210}
\address{$^2$Physics Department, McGill University, 3600 University Street,
Montr\'eal, Canada H3A 2T8}
\maketitle

\begin{abstract}
\noindent
We revisit the emissivity from neutrino pair bremsstrahlung in 
neutron-neutron scattering, $n n \rightarrow n n \, \nu
{\bar{\nu}}$, which was calculated from the one-pion exchange 
potential including correlation effects by Friman and Maxwell. 
Starting from the free-space low-momentum nucleon-nucleon 
interaction $\vlk$, we include tensor, spin-orbit and
second-order medium-induced non-central contributions to the 
scattering amplitude in neutron matter. We find that the screening 
of the nucleon-nucleon interaction reduces the emissivity from 
neutrino bremsstrahlung for densities below nuclear 
matter density. We discuss the implications of medium modifications 
for the cooling of neutron stars via neutrino emission, taking into 
account recent results for the polarization effects on neutron 
superfluidity.
\end{abstract}

\section{Introduction}

The observation of neutron star properties and their evolution
provides a challenging astrophysical setting for the study of dense
nuclear matter~\cite{nstar1,nstar2}. On the theoretical side, the main
objectives are a more comprehensive understanding of
the equation of state and of microscopic nuclear properties, such as
pairing and transport phenomena, where progress is tied to improving
many-body calculations and techniques for densities ranging from
sub-nuclear to a few times saturation density. Accordingly, and spurred 
by the study of rare isotopes, future theoretical research is also
directed towards a systematic study of neutron-rich matter.

A unique method to probe the internal composition of a
neutron star is by tracing its temperature evolution with cooling
simulations, see e.g.,~\cite{chris,page,cooling1,cooling2}. Neutron
stars are born in supernova explosions and the interior temperatures 
initially exceed $T \sim 10^{11} \, \text{K}$. As the neutron star 
cools, neutrinos begin to free stream
and essentially leave the star without further interaction. As a
consequence, after about $30 \, \text{s}$, the long-term cooling of
neutron stars is controlled by neutrino emission. This stage lasts
up to about $10^5$ years of age, when cooling by emission of photons
becomes more effective. Importantly, the neutron stars remain luminous
enough during the cooling, so that the surface temperature can be
extracted from space telescope data and the theoretical cooling curves
can be confronted with observations.

For proton fractions $n_p/(n_n+n_p) < 1/9$, direct beta decay does 
not proceed in neutron star matter due to the imbalance of the neutron 
and proton Fermi momenta. Therefore, and in the absence of nucleon 
superfluidity, the dominant neutrino
emission comes from $\nu {\bar{\nu}}$ bremsstrahlung in
nucleon-nucleon collisions and the so-called modified Urca
process. The latter corresponds to in-medium beta decay, where the
momentum difference between the decaying neutron and the final proton
is absorbed by scattering off a second nucleon. For low temperatures
$T \ll E_{F_{n,p}}$ (where $E_F$ denotes the Fermi energy and 
we use $k_B = 1$),
the temperature dependence of these different emission channels is
easily power-counted via the degeneracy of the fermions involved in
the emission. It follows that the emissivity $\varepsilon$, which is the
total neutrino energy emitted per unit volume and unit time, scales as
$\varepsilon \sim T^8$.

However, the density dependence and strength of the emissivity requires,
as input, accurate calculations of the nucleon-nucleon scattering 
amplitude in the many-body medium. Specifically, the bremsstrahlung 
processes involve the scattering amplitude for nucleons on the Fermi 
surface, whereas the modified Urca process involves also off-shell 
scattering, when e.g., the scattered neutron beta decays and thus 
its momentum is approximately given by the proton Fermi momentum.
A further theoretical challenge lies in the fact that the dominant 
contribution to neutrino emissivities comes (for bremsstrahlung
in $nn$ collisions 
exclusively) from the non-central parts of the nucleon-nucleon amplitude, 
in particular from the tensor force~\cite{FM}. In addition to neutrino 
emissivities, the general importance of non-central interactions for 
neutron star properties has been revived recently, where the effects 
on the magnetic susceptibility~\cite{OP} as well as on P-wave pairing
in neutron star cores~\cite{AB} have been demonstrated to be crucial.

For the nucleon scattering amplitude, the benchmark calculation of 
Friman and Maxwell takes into account the long-range one-pion 
exchange tensor force explicitly and estimates the effects 
of short-range correlations by cutting off the interaction at short 
distances and including the short-range rho-exchange tensor 
force~\cite{FM}. In this work, we start from the free-space
low-momentum nucleon-nucleon interaction $\vlk$~\cite{Vlowk1,Vlowk2}.
The construction of $\vlk$ is motivated by the differences of 
the realistic nucleon-nucleon potential models at short distances. 
The model dependence at short distances $r < d \approx 0.5 \, \text{fm}$ 
originates from the fact that the interaction cannot be resolved 
from scattering experiments probing low momenta $p < \Lambda \approx 
2.0 \, \text{fm}^{-1}$ (where $\Lambda = 1/d$). Note that the realistic
potential models are fitted to phase shifts below $E_\text{lab} \approx 350
\, \text{MeV}$, corresponding to $\Lambda \approx 2.1 \, \text{fm}^{-1}$.
A systematic method to remove the model dependence 
is provided by the renormalization group (RG), where the high momentum 
modes with $p \geqslant \Lambda$ are integrated out to construct the
physically equivalent effective theory. The renormalization group 
in this context is used as a tool to guarantee that the phase shifts 
are preserved by the low-momentum interaction under the 
renormalization. Since $\vlk$ does not have momentum components larger
than the cutoff $\Lambda$, it does not have a strongly repulsive core.
As a consequence, one does not have to compute a Brueckner $G$ matrix 
from $\vlk$ in many-body applications. In the sense of the RG, short-range 
correlation effects are implicitly included in $\vlk$~\cite{IIpaper}.

In addition to the phenomenological short-range correlation effects 
discussed by Friman and Maxwell, the scattering amplitude is screened
by the particle-hole polarization of the many-body medium. For the 
in-medium tensor force, particle-hole screening effects are very
important. This follows from a general spin-recoupling argument 
due to the interference of the central spin-spin part and the tensor 
part of the nucleon-nucleon interaction, leading to a substantial 
decrease of the latter~\cite{AB,IIpaper}. Moreover, the presence of the
Fermi sea defines a preferred frame, which leads to novel non-central
parts in the effective interaction and the scattering amplitude in the
many-body medium~\cite{AB,Forest}. In pure neutron matter, the scattering 
amplitude on the Fermi surface has been computed to second-order in 
$\vlk$, with particular attention to the spin-dependence and non-central
interactions, where it was found that the particle-hole screening leads to
a substantial decrease of the tensor force and a significant 
long-wavelength center-of-mass tensor force induced by the 
medium~\cite{AB}. As our understanding of the renormalization of the
nucleon-nucleon interaction in dense matter improves, it is thus 
important to include the in-medium modifications of the nuclear force
as well as the novel non-central contributions in the calculation
of the neutrino emissivities. Furthermore, with recent results for
the neutron superfluid S- and P-wave pairing gaps including polarization 
effects obtained in~\cite{AB,RGnm}, a consistent calculation of the
emissivity is needed.

The neutrino emissivities also receive contributions from the 
spin-orbit force, which were included in recent calculations of
Hanhart {\it et al.}~\cite{Hanhart} and van Dalen {\it et 
al.}~\cite{Dalen} starting from the free-space scattering
amplitude. As for the tensor force, particle-hole polarization 
effects reduce the spin-orbit interaction in neutron matter~\cite{AB}. 
As a consequence, the $^3$P$_2$ superfluid pairing gaps in neutron
star cores are strongly suppressed to below few $\text{keV}$
at second order in $\vlk$ for the pairing interaction~\cite{AB}.
Therefore, one expects that the transition to the P-wave superfluid
phase of neutrons is only reached in the very late stages of neutron 
star cooling. The consistency with data in present cooling simulations
of Yakovlev {\it et al.} also requires low critical temperatures of
the $^3$P$_2$ superfluid, $T_c < 2 \cdot 10^8 
\, \text{K}$~\cite{cooling2}. This corresponds
to an angle-averaged gap (as calculated in~\cite{AB}) in the $m_J=0$ state
of $\Delta < 30 \, \text{keV}$. We note that on the level of the
free-space nucleon-nucleon interaction considerably larger 
P-wave pairing gaps of $\Delta \approx 0.35 \, \text{MeV}$ are 
predicted at nuclear matter density~\cite{Baldo}.

If we thus consider neutron star matter 
above nuclear matter density, the 
$^1$S$_0$ superfluidity of neutrons ceases to exist due to the 
repulsion in the nuclear force and one expects that, at the relevant
core temperatures, the neutrons are in the normal phase and the protons 
are superconducting. In this case, the modified Urca process is strongly 
suppressed due to proton superfluidity, and the dominant neutrino emission 
process will be neutrino bremsstrahlung in $nn$ collisions. This is the 
motivation to focus on the in-medium modification of the bremsstrahlung 
rate in this work. The second motivation comes from the fact that 
even at lower densities, where both neutrons and protons pair in the
$^1$S$_0$ channel, the neutrino pair emissivity is more effective compared to
the modified Urca process~\cite{Yakovlev}, although this conclusion depends
on the values of the pairing gaps and neutrino emission mainly proceeds 
through the so-called Cooper Pair-Breaking and Formation (PBF) process in 
the regime $0.2 \, T_c \lesssim T < T_c$~\cite{Flowers}.

In this work, we use as input the second-order results for the 
neutron-neutron scattering amplitude on the Fermi surface computed 
in~\cite{AB}. We start by giving the general spin-dependence of the
scattering amplitude on the Fermi surface, with a short discussion of
the novel contributions in Section~\ref{spindep}. The non-central
interactions are included in the calculation of the emissivity from 
neutrino bremsstrahlung in
Section~\ref{emiss}. This Section follows closely the derivation
of Friman and Maxwell~\cite{FM}. The results for the emissivity over a 
range of densities is given at the end of Section~\ref{emiss}.
Finally, we conclude with a discussion of superfluidity in neutron
stars, which takes into account recent results for the pairing
gaps~\cite{AB,RGnm}. We present some general arguments for the
constraints on the density dependence of the gaps and compare the
bremsstrahlung rate to the PBF process for various temperatures. 
Revised estimates of the $np$ and $pp$ bremsstrahlung and 
modified Urca rates will be reported 
in a subsequent publication, as they require the nucleon-nucleon
scattering amplitude off the Fermi surface as well as an extension
to asymmetric matter.

\section{Spin-Dependence of the Scattering Amplitude in Neutron Matter}
\label{spindep}

The scattering amplitude for neutrons on the Fermi surface contains the
free-space central (scalar and spin-spin) and non-central (spin-orbit 
and tensor) parts. In addition, the many-body medium can induce effective 
interactions, which depend on the two-body center of mass momentum 
$\Pt={\bm p}_1+{\bm p}_2={\bm p}_3+{\bm p}_4$ and one has in general~\cite{AB}
\beqy
{\mathcal{A}}_{\si, \sip}(\q,\qp,\Pt) &=& \sum_i  
{\mathcal{A}}_i (q,q') \, {\mathcal{O}}^i_{\si, \sip}(\qh,\qph,\Pth) \\[1mm]
&=& {\mathcal{A}}_{\text{scalar}}(q,q') 
+ {\mathcal{A}}_{\text{spin}}(q,q') \, \si \cdot \sip
+ {\mathcal{A}}_{\text{spin-orbit}}(q,q') \, i (\si + \sip) 
\cdot \qh \times \qph \nonumber \\[1mm]
&+& {\mathcal{A}}_{\text{tensor}}(q,q') \, S_{12}(\qh)
+ {\mathcal{A}}_{\text{exch. tensor}}(q,q') \, S_{12}(\qph)
+ {\mathcal{A}}_{\text{cm tensor}}(q,q') \, S_{12}(\Pth) \nonumber \\[1mm]
&+& {\mathcal{A}}_{\text{diff. vector}}(q,q') \,
i (\si - \sip) \cdot \qh \times \Pth
+ {\mathcal{A}}_{\text{cross vector}}(q,q') \,
(\si \times \sip) \cdot (\qph \times \Pth) ,
\label{amp}
\eeqy
where $\q={\bm p}_1-{\bm p}_3$ and $\qp={\bm p}_1-{\bm p}_4$ denote the 
momentum transfers in the direct and exchange channel respectively,
the tensor operator is defined as $S_{12}(\qh) \equiv \si \cdot 
\qh \, \sip \cdot \qh - 1/3 \, \si \cdot \sip$, and the scattering
amplitude is defined in units of the density of states $m_n^\ast \kfn/\pi^2$.
For particles on the Fermi surface, the momentum transfers and the
center-of-mass momentum are orthogonal and one has $q^2+q'^2+P^2=4 \kfns$.
Therefore, the various parts ${\mathcal{A}}_i$ depend only on the magnitude
of the momentum transfers $q=|\q|$ and $q'=|\qp|$. We also note that the 
non-central operators are defined with unit vectors~\cite{OP}. Finally,
the tensor operators given in Eq.~(\ref{amp}) are linearly dependent, with 
$S_{12}(\qh) + S_{12}(\qph) + S_{12}(\Pth)=0$. In the second-order 
calculation all tensors are kept explicitly~\cite{AB}, and for the 
emissivities we then eliminate $S_{12}(\Pth)$, leading 
to $\widetilde{\mathcal{A}}_\text{tensor}={\mathcal{A}}_\text{tensor}
-{\mathcal{A}}_\text{cm tensor}$ as well as
$\widetilde{\mathcal{A}}_\text{exch. tensor}={\mathcal{A}}_\text{exch. tensor}
-{\mathcal{A}}_\text{cm tensor}$.
In the scattering amplitude, both direct and exchange terms are accounted for; 
e.g., to lowest order, one has ${\mathcal{A}} = \vlk - P_{\bm \sigma}
P_{\bm k} \vlk$, with spin- and momentum-exchange operators $P_{\bm \sigma}$
and $P_{\bm k}$. The latter two operators in Eq.~(\ref{amp}) do not 
conserve the spin of the interacting particle pair and are induced in 
the medium due to the screening by particle-hole excitations. 

Although it is generally known that particle-hole polarization
effects are very 
important in nuclear physics, this is the first calculation where 
these are included for neutrino emissivities. In the following
we compute the emissivity without superfluid effects, in order
to assess the renormalization of the tensor and spin-orbit forces in
matter and the contributions of the novel non-central forces, where
the cm tensor is included in the conventional tensor parts. Subsequently, 
we discuss polarization effects for the superfluid properties and 
compare the bremsstrahlung rate to the dominant PBF process~\cite{Flowers}. 
In both parts we give results for the density dependence of the emissivity.

\begin{figure}[t]
\begin{center}
\includegraphics[scale=1,clip=]{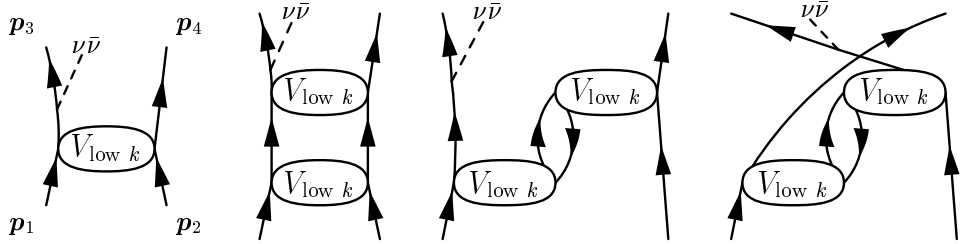}
\end{center}
\caption{Feynman diagrams contributing to the neutrino emissivity
from bremsstrahlung to second-order in $\vlk$. The $\vlk$ vertex includes
both the direct and the exchange term and the dashed line corresponds
to the neutral current. As in~[7], the emissivity includes
permutations of the neutral current attached to all external lines, 
whereas the coupling to the internal nucleon lines is suppressed,
because it does not lead to the small energy denominator in the 
additional nucleon propagator as in Eq.~(5).}
\label{Feyndiag}
\end{figure}

\section{Neutrino Bremsstrahlung in Neutron-Neutron Collisions}
\label{emiss}

Following Friman and Maxwell~\cite{FM}, the emissivity from neutrino 
pair bremsstrahlung in neutron-neutron scattering is given by 
(for $\hbar=c=1$)
\beq
\label{emissmain}
\varepsilon_{nn} = N_{\nu} \int \biggl( \, \prod_{i=1}^4
\frac{d^3 {\bm p}_i}{(2\pi)^3} \biggr) \,
\frac{d^3 {\bm Q}_1}{2\omega_1(2\pi)^3} \,
\frac{d^3 {\bm Q}_2}{2\omega_2(2\pi)^3} 
\, (2 \pi)^4 \, \delta(E_f-E_i) \, \delta^3({\bm P}_f - {\bm P}_i) 
\, \frac{1}{s} \, \biggl( \: \sum_{\text{spin}} \, 
\bigl| \, {\mathcal{M}}_{nn} \bigr|^2 \biggr) 
\, \omega_{\nu} \, {\mathcal{F}}(E_{{\bm p}_i}) ,
\eeq
where ${\bm p}_i$ denote the momenta of the incoming and outgoing
neutrons and $Q_{1,2}=(\omega_{1,2},{\bm Q}_{1,2})$ 
label the neutrino energies and
momenta. The delta functions account for energy and momentum conservation, 
and $\omega_{\nu}=\omega_1+\omega_2$ is the total neutrino energy.
$N_{\nu}$ denotes the number of neutrino species and $s=2$ is a 
symmetry factor for the initial neutrons, when the emission occurs in
the final state or vice versa.
The function ${\cal{F}}(E_{{\bm p}_i})=f(E_{{\bm p}_1}) \, 
f(E_{{\bm p}_2}) \, (1-f(E_{{\bm p}_3})) \, (1-f(E_{{\bm p}_4}))$ 
is the product of Fermi-Dirac distribution functions
$f(E)=(\exp(E/T) + 1)^{-1}$, with neutron energies $E_{{\bm p}_i}$.
The matrix element ${\mathcal{M}}_{nn}$ includes the nucleon-nucleon
scattering part and the coupling to the emitted neutrino pair.
For the bremsstrahlung process the corresponding Feynman diagrams
at tree-level and with second-order contributions are shown in
Fig.~\ref{Feyndiag}. The second-order particle-hole intermediate 
states include all possible excitations for interacting particles 
on the Fermi surface. In the particle-particle channel, the cutoff 
in $\vlk$ provides a regulator, and we evaluate the phase space 
(including hole-hole states) exactly without angle-averaging 
approximation. We note that the logarithmic contribution to the 
quasiparticle interaction (due to the BCS singularity) is integrable
and leads to finite Fermi liquid parameters~\cite{Sjoberg}, and 
thus finite emissivities.

Neutrino pair emission from a neutron line is given by the neutral 
current $V-A$ weak interaction
\beq
{\mathcal{L}}^n_\text{neutral} = -\frac{G_F}{2\sqrt{2}} \, \chi_1^{\dag}
\, ( \delta_{\mu 0} - g_A \, \delta_{\mu i} \, \sigma_i ) \, \chi_2 \,
l_\mu ,
\eeq
with Fermi coupling constant $G_F=1.166 \cdot 10^{-5} \, \text{GeV}^{-2}$, 
the neutrino current $l_\mu = \bar{u}(Q_1) \, \gamma_\mu \, (1-\gamma_5)
\, u(Q_2)$ and the weak axial-vector coupling constant $g_A=1.26$.
The non-relativistic nucleon spinors are denoted by $\chi$ and $u(Q_{1,2})$ 
are relativistic spinors for the neutrinos, which are taken to be massless. 
As in~\cite{FM}, we use a non-relativistic approximation for all nucleon 
propagators, where the lowest term in an expansion in inverse powers of 
the nucleon mass is retained. In addition, one neglects the neutrino 
pair energy $\omega_\nu$ compared to the Fermi energy, since the emitted 
neutrinos are thermal. For the nucleon propagator $G$, this approximation 
yields
\beq 
i \, G({\bm p} \pm {\bm Q}_\nu,E_{\bm p} 
\pm \omega_\nu) = \pm i \, \omega_\nu^{-1} ,
\label{smallomega}
\eeq 
where the positive sign holds if the weak current is attached to an
outgoing nucleon, negative otherwise. It follows that, in the non-relativistic 
approximation for the nucleon propagators, the vectorial part of the
neutral current does not contribute~\cite{FM}.

The matrix element ${\mathcal{M}}_{nn}$ includes the strong interaction 
part with spin-dependence given by Eq.~(\ref{amp}). The spin sum over
the squared matrix element is carried out independently for the pieces
coming from the nucleon-nucleon amplitude and the weak interaction, where
only the non-central parts in the amplitude are found to contribute. After
contraction with the lepton trace given by 
\beq 
{\rm Tr} \bigl( l_i \, l_j^{\dag} \bigr) = 8 \, \bigl( 
{\bm Q}_{1_i} \, {\bm Q}_{2_j}
+ {\bm Q}_{2_i} \, {\bm Q}_{1_j} - g_{ij} \, Q_1 \cdot Q_2
+ i \, \epsilon_{i\alpha j\beta} \, Q_1^{\alpha} \, Q_2^{\beta} \bigr) ,
\eeq
one finds
\beqy
\sum_\text{spin} \, \bigl| \, {\mathcal{M}}_{nn} \bigr|^2
&=& 64 \, g_A^2 \, G_F^2 \, \frac{\omega_1 \, \omega_2}{\omega_\nu^2}
\, \frac{\pi^4}{m_n^{\ast\,2} \kfns} \, {\mathcal{A}}^2_{nn}(q,q')
\nonumber \\[1mm]
&=& 64 \, g_A^2 \, G_F^2 \, \frac{\omega_1 \, \omega_2}{\omega_\nu^2}
\, \frac{\pi^4}{m_n^{\ast\,2} \kfns}
\biggl( \widetilde{\mathcal{A}}^{\,2}_{\text{tensor}}(q,q')
+ \widetilde{\mathcal{A}}^{\,2}_{\text{exch. tensor}}(q,q')
- \widetilde{\mathcal{A}}_{\text{tensor}}(q,q')
\, \widetilde{\mathcal{A}}_{\text{exch. tensor}}(q,q') \nonumber \\[1mm]
&+& {\mathcal{A}}^{\,2}_{\text{spin-orbit}}(q,q')
+ {\mathcal{A}}^{\,2}_{\text{diff. vector}}(q,q')
+ 3 \, {\mathcal{A}}^{\,2}_{\text{cross vector}}(q,q') \biggr) ,
\label{nnmatrix}
\eeqy
where we have dropped terms that vanish upon angular integrations
over ${\bm Q}_{1,2}$ in the emissivity, Eq.~(\ref{emissmain}), when 
the neutrino momenta are neglected in the momentum-conserving delta
function. 

\begin{figure}[t]
\begin{center}
\includegraphics[scale=0.45,clip=]{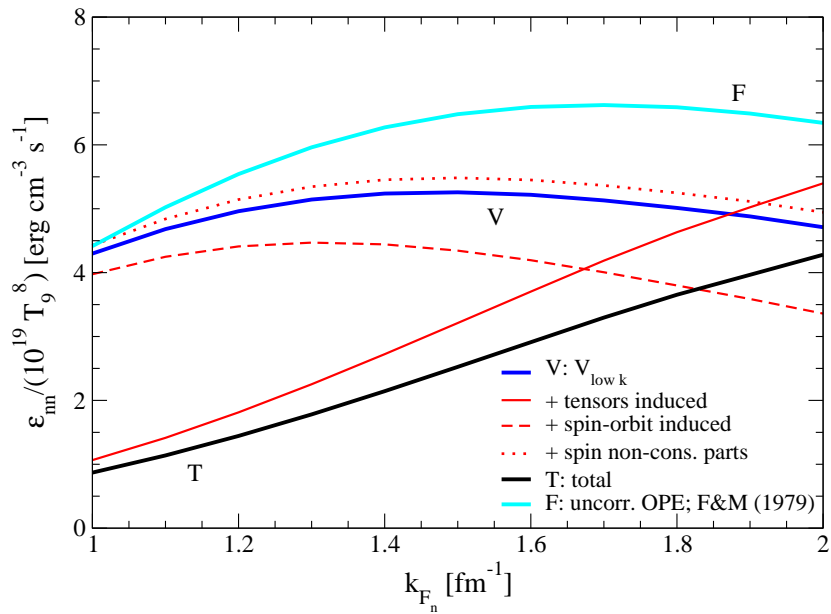}
\end{center}
\caption{The neutrino emissivity from bremsstrahlung in neutron-neutron 
scattering $\varepsilon_{nn}$ versus Fermi momentum $\kfn$ in neutron
matter. All curves include the lowest-order effective mass obtained from 
$\vlk$, see~[9]. The curve labeled V denotes the lowest-order emissivity 
obtained from the free-space low-momentum interaction. In comparison, 
we give the results obtained by Friman and Maxwell (curve F) using the 
uncorrelated one-pion exchange (OPE) tensor force without the exchange 
terms [7]. (In [7], the inclusion of exchange terms, the $\rho$ tensor 
force, and short-range correlation effects yields a multiplicative
suppression factor of $\approx 0.62$ at nuclear matter density or
$\kfn = 1.7 \, \text{fm}^{-1}$.) The thin curves correspond to 
the $\vlk$ result with,
respectively, second-order renormalization of the tensor, spin-orbit
or spin non-conserving forces in the medium included. The curve labeled T 
is the full second-order result, where both particle-hole channels and the 
particle-particle channel are taken into account.}
\label{nnemiss}
\end{figure}

For the evaluation of the phase space, one can easily perform the 
integrals over the neutrino momenta with $|{\bm Q}_{1,2}| = \omega_{1,2}$ 
by inserting
\beq 
1 = \int d \omega_\nu \, \delta(\omega_\nu-\omega_1-\omega_2) ,
\eeq
as the energy-conserving delta function depends only on the total
neutrino energy. Moreover, one can decouple the angular parts in the
neutron phase space and trade the radial momentum for energy integrals,
by restricting the interacting neutrons to the Fermi surface, since they 
are strongly degenerate for typical neutron star temperatures. Corrections
to this approximation scale as $T/E_{F_n}$. For this
purpose, one replaces
\beq 
d^3 {\bm p}_i \rightarrow d^3 {\bm p}_i \, \frac{m_n^\ast}{\kfn}
\, \delta(p_i - \kfn) \, \int dE_{{\bm p}_i} .
\eeq 
Finally, we introduce the integration over momentum transfers through
respective delta functions in the direct and the exchange channels,
\beq
1 = \int d^3\q \, \delta^3(\q - {\bm p}_1 + {\bm p}_3)
= \int d^3 \qp \, \delta^3(\qp - {\bm p}_1 + {\bm p}_4) .
\eeq
After carrying out the angular integrations, we find the general expression 
for the emissivity
\beq
\varepsilon_{nn} = \frac{64}{15} \frac{g_A^2 \, G_F^2 \, 
m_n^{\ast\,2}}{2^9 \, \pi^6 \, \kfn} \, N_{\nu} \, I_{\nu \bar{\nu}}
\, \langle {\mathcal{A}}^2_{nn} \rangle = 
0.781 \, T^8_9 \, N_{\nu} 
\, \biggl(\frac{m_n^\ast}{m_n}\biggr)^2
\, \biggl(\frac{1.7 \, \text{fm}^{-1}}{\kfn}\biggr) \: 
\langle {\mathcal{A}}^2_{nn} \rangle 
\: \text{erg cm}^{-3} \, \text{s}^{-1} ,
\eeq
where $I_{\nu\bar{\nu}}$ is the result of the integrals over the 
total neutrino and neutron 
energies convoluted with the thermal distribution factors 
${\mathcal{F}}(E_{{\bm p}_i})$ which is identical 
to the expression in~\cite{FM},
and the square of the scattering amplitude averaged over the Fermi surface 
$\langle {\mathcal{A}}^2_{nn} \rangle$ is given by
\beq
\langle {\mathcal{A}}^2_{nn} \rangle =
\int\limits_0^{2 \kfn} \frac{dq}{\kfn} \, \int\limits_0^{2 \kfn}
\frac{dq'}{\kfn} \, \frac{\kfn \, \Theta(4 \kfns - q^2 - 
q^{\prime\,2})}{\bigl(4 \kfns - q^2 - q^{\prime\,2}\bigr)^{1/2}} \:
{\mathcal{A}}^2_{nn}(q,q')
= \int\limits_0^{2 \kfn} \frac{dq}{\kfn} \, \int\limits_0^{\pi/2}
d\phi \: {\mathcal{A}}^2_{nn}\bigl(q,\sqrt{4 \kfns - q^2} \, \sin\phi
\bigr) .
\label{emiss1}
\eeq
The remaining two-dimensional integral is calculated numerically and our 
results for $N_{\nu}=3$ neutrino flavors are shown in Fig.~\ref{nnemiss}
for densities ranging from $\kfn = 1.0 - 2.0 \, \text{fm}^{-1}$. All results
given in Fig.~\ref{nnemiss} include the effective mass obtained from
the lowest order $\vlk$. The effective mass varies from $m^\ast_n/m_n = 
0.95$ at $\kfn = 1.0 \, \text{fm}^{-1}$ to $m^\ast_n/m_n = 0.78$ at $\kfn 
= 2.0 \, \text{fm}^{-1}$, and is in this range well approximated by a
linear curve versus Fermi momentum. We note that one expects an increase 
of the effective mass in the induced interaction (which is compensated
by the quasiparticle strength $z_{\kf}$), as can be seen from the 
results of the full RG calculation for neutron matter~\cite{RGnm}.
Furthermore, our results do not include a renormalization of $g_A$ 
in the medium to $g_A \approx 1.0$.

First, we compare the lowest-order $\vlk$ results to the
calculation of Friman and Maxwell from uncorrelated one-pion
exchange (OPE), see Eq.~(52) in~\cite{FM}. We remark that exchange 
terms, the inclusion of the $\rho$ tensor force and correlation
effects lead to a multiplicative suppression factor in the calculation of 
Friman and Maxwell of $\approx 0.62$ at nuclear matter density. We 
find that $\vlk$ gives 
similar rates, but without the need to estimate the correlation 
distance, which is experimentally unconstrained in neutron matter.

Recently, Hanhart {\it et al.}~\cite{Hanhart} and van Dalen {\it
et al.}~\cite{Dalen} also computed the neutrino pair emissivity from
bremsstrahlung employing Low's theorem for soft emission, with the
free-space on-shell $T$ matrix as input. These results provide a 
model-independent low-density limit on the emissivity from neutrino 
bremsstrahlung. In both works, the emissivity was found to be reduced 
by a factor $\approx 1/4$ compared to the OPE result of Friman and 
Maxwell at saturation density. However, the applicability to relevant 
neutron star densities is limited when one starts from the free-space 
scattering amplitude. This is due to the fact that at 
second-order in the T matrix the spin-spin and tensor part of the 
amplitude mix due to screening in the particle-hole channel. If one 
denotes the tensor part of the free-space amplitude by $T_\text{tensor}$, then 
the second-order contribution will be proportional to $T_\text{tensor}
\, \kfn \, a_\text{S}$ at low momenta, where $a_\text{S}$ is the 
S-wave scattering length coming from the spin-spin part of the
$T$ matrix. Even at low-densities, e.g., $\kfn = 1/2 \, \text{fm}^{-1}$ 
(i.e., $\rho = 1/40 \, \rho_0$), this would be a very large correction, 
which is not accounted for in~\cite{Hanhart,Dalen}.

Beyond the lowest-order result, we find that the renormalization
of the non-central parts of the nucleon-nucleon interaction in the
medium considerably reduces the emissivity, especially at sub-nuclear 
densities, as shown in Fig.~\ref{nnemiss}. We emphasize
that a second-order calculation cannot give final results, but it
provides a range for the effects due to particle-hole screening.
From the denominators of the induced interaction~\cite{BB}, one expects 
higher orders to somewhat decrease the second-order results. However, 
this argument is not straightforward for non-central interactions and 
a definitive conclusion requires explicit calculations of higher-orders 
with full non-central spin-dependence. Referring to a previous calculation
of the induced quasiparticle interaction (i.e., for $q = 0$) including 
tensor forces~\cite{Dickhoff}, it was also found that the tensor force 
is reduced in the medium. For the cooling of neutron stars, as well as 
spin-isospin response in supernovae, the modification of transport 
properties in the medium is very important. Our work shows that 
particle-hole effects must be taken into account in a realistic 
calculation of the emissivities.

As discussed in the Introduction, an accurate estimate of the bremsstrahlung 
rate is important particularly at high densities, since the competing 
processes except for possible emission by PBF are expected to be 
strongly suppressed due to proton superfluidity. Therefore, we will 
compare the strength of the open PBF channel to the emission from a 
non-superfluid core of neutrons in the next Section.

\section{Superfluidity and Comparison with PBF processes}
\label{superfluid}

The temperatures in the interior of cooling neutron stars can be well 
below the critical temperature for neutron or proton superfluidity.
This leads to a strong reduction of the neutrino emissivities since 
the fraction of particles that are unpaired scales exponentially
with the temperature as $\exp(-2 \Delta/T)$, where $\Delta$ denotes the
zero temperature gap. Thus, the modification of the emissivities relies 
on an accurate calculation of the proton and neutron pairing gaps in 
neutron star matter, which must include the particle-hole polarization 
in the medium.

Before discussing recent results for the neutron pairing gaps in pure
neutron matter~\cite{AB,RGnm}, we proceed with some remarks on
the superfluid properties and rather general in-medium modification
arguments. The strongest attraction in the nuclear force 
is in the S-wave for laboratory energies below $E_{\text{lab}} \lesssim 
250 - 260 \, \text{MeV}$, i.e., for back-to-back scattering of particles
with momenta $k = \kfn \lesssim 1.7 - 1.8 \, \text{fm}^{-1}$. Thus, for
densities below nuclear matter density $\rho \lesssim \rho_0$, 
one expects that both neutrons and protons form a superfluid in 
the isotriplet $^1$S$_0$ channel (Note that for typical proton 
fractions of $n_p/(n_n+n_p) \approx 0.05$, one has for the proton 
Fermi momentum $\kfp \approx 1/3 \, \kfn$). At higher
densities, one concludes from the free-space scattering phase shifts 
that the neutrons are expected to pair in the 
$^3$P$_2$ state~\cite{Hoffberg}. Eventually,
the ground state of matter at high densities has to be determined in a
model of dense matter, as realistic interactions are constrained to
relative momenta $k \lesssim 2.1 \, \text{fm}^{-1}$.

It is well known that polarization effects on the nucleon-nucleon
interaction, which are necessary in order to satisfy the Pauli 
principle in microscopic calculations~\cite{BB}, lead to a strong 
reduction of the superfluid gaps in neutron 
stars~\cite{AB,RGnm,Clark,WAP,3pf2}. The effect of the induced 
interaction on pairing can be understood by considering 
the second-order contributions. Higher-order terms modify 
the strength of the second-order result, but usually do not 
alter whether the induced interaction is attractive or repulsive 
in the particular channel.

For S-wave pairing of neutrons, the dominant part in the quasiparticle
interaction is the central spin-spin delta function $G_0 \approx 
0.6-0.8$~\cite{RGnm}. Neglecting the smaller contributions, one has
for the induced pairing interaction in the $S=0$ state,
\beq
{\mathcal A}_{\text{central}}^{\text{ind}}(\cos\theta_\q) = 3 \, G_0^2 \, 
\bigl( U(q/\kfn) + U(q'/\kfn) \bigr) ,
\label{centralind}
\eeq
where direct and exchange particle-hole channels are accounted for,
$U(q/\kfn)$ denotes the (positive) static Lindhard function and for
back-to-back scattering $q = \kfn \sqrt{2 - 2\cos\theta_\q}$ 
($q' = \kfn \sqrt{2 + 2\cos\theta_\q}$) with scattering angle 
$\theta_\q$. The projection of the sum of Lindhard 
functions in Eq.~(\ref{centralind}) on S-wave yields $\langle U(q/\kfn) 
+ U(q'/\kfn) \rangle_{l=0} = 2 (1+2\log2) / 3 \approx 1.59$ (see 
also~\cite{heiselberg}). As a result, spin fluctuations will reduce 
the pairing interaction and consequently the superfluid gap will close
at somewhat lower densities than 
one would expect from the free-space $^1$S$_0$ 
phase shifts. In the RG calculation of the effective interaction and the
scattering amplitude on the Fermi surface~\cite{RGnm} (for a discussion 
of the RG approach see also~\cite{Hirschegg}), it is found that
the maximum $^1$S$_0$ pairing gap is reduced to $0.8 \, \text{MeV}$ at
$\kfn \approx 0.8 \, \text{fm}^{-1}$ and that the gap disappears at 
$\kfn \approx 1.5 \, \text{fm}^{-1}$ or $\rho \approx 2/3 \, \rho_0$. 
This corresponds to a maximum critical temperature of $T_c = 5.3 \cdot
10^9 \, \text{K}$.

The study of polarization effects on P-wave pairing at higher 
densities is more involved, since spin-orbit and tensor forces are 
crucial for the reproduction of the P-wave phase shifts in vacuum and
consequently one has to address the renormalization of non-central 
interactions in the medium. In particular, the medium-induced 
spin-orbit force leads to a strong suppression of the $^3$P$_2$ gaps, 
which is e.g., due to the interference of the central spin-spin and 
the spin-orbit force at second order~\cite{AB}. A similar argument as 
for S-wave pairing reproduces the effect of induced pairing interactions 
qualitatively and one has
\beq
{\mathcal A}_{\text{spin-orbit}}^{\text{ind}}(\cos\theta_\q) 
= - \frac{1}{2} \, G_0 \, \langle V_{\text{SO}} \rangle \, 
\bigl( U(q/\kfn) + U(q'/\kfn) \bigr) \, \frac{q q'}{\kfns} ,
\label{soind}
\eeq
where the dimensionless $\langle V_{\text{SO}} \rangle < 0$ denotes an 
averaged spin-orbit matrix element. The induced contributions 
due to the mixing of spin-orbit and tensor forces are also repulsive,
with a similar but more complicated momentum dependence. For the
induced $^3$P$_2$ pairing matrix element, the largest contribution comes from
the same $l=0$ projection of the sum of the Lindhard functions, where
the additional momentum-dependent factors $q q'$ are absorbed in the 
spin-orbit operator ${\bm L} \cdot {\bm S} \sim i (\si + \sip) \cdot 
\q \times \qp$ (note the unit vectors in Eq.~(\ref{amp})). In~\cite{AB}
it was found that second-order polarization effects lead to a strong
depletion of the $^3$P$_2$ pairing gap from $\Delta \approx 0.3 \, 
\text{MeV}$ (obtained from the free-space $\vlk$) to superfluid
gaps on the level of few $\text{keV}$ at nuclear matter density. 
This corresponds to a ratio of $0.45$ of the second-order to lowest-order 
pairing interaction, and therefore it is expected that the suppression
of the gap at second order is robust with pairing gaps below $\Delta 
\lesssim 1-10 \, \text{keV}$. The latter value corresponds to a critical 
temperature of $T_c = 10^{6.8-7.8} \, \text{K}$. With surface temperatures 
from observational data $T_\text{s}^\infty \geqslant 10^{5.6} \, 
\text{K}$, i.e., core temperatures $T \geqslant 10^{7.4} \, \text{K}$
(see e.g., Table~2 in~\cite{Shibanov99}), 
it is thus expected that the $^3$P$_2$
superfluid phase is only reached at late cooling stages. In fact, 
Yakovlev {\it et al.} have checked that for critical temperatures $T_c 
< 2 \cdot 10^8 \, \text{K}$ the $^3$P$_2$ phase has no impact for the
cooling of middle-aged neutron stars~\cite{cooling2}. We therefore
proceed and compare the bremsstrahlung emissivity in normal matter
to the emission of neutrinos from the $^1$S$_0$ superfluid condensate
only.

\begin{figure}[t]
\begin{center}
\includegraphics[scale=0.45,clip=]{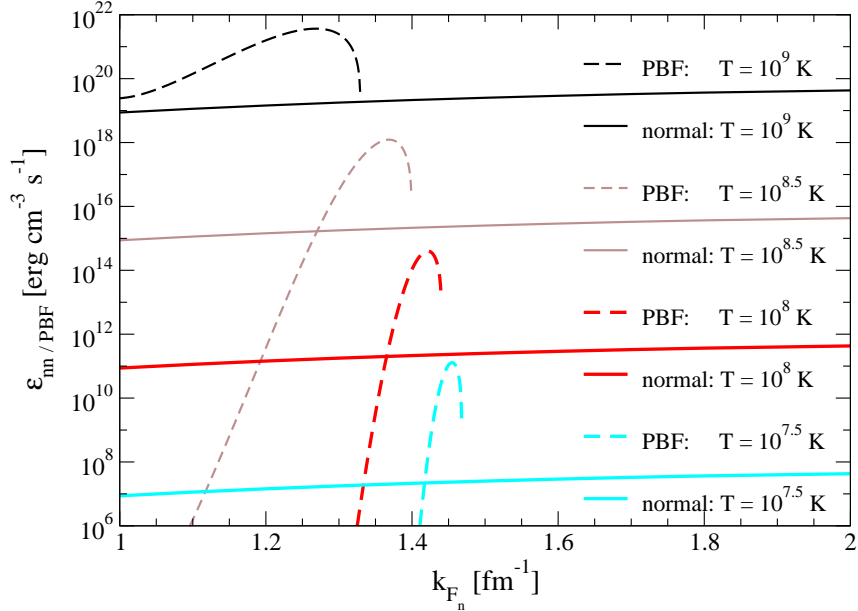}
\end{center}
\caption{Comparison of the emissivity from neutrino pair bremsstrahlung
in non-superfluid neutron matter (total in Fig.~2) and the $^1$S$_0$
PBF process as a function of Fermi momentum and for various temperatures. 
Results are shown using the $^1$S$_0$ superfluid gaps obtained in the 
RG approach with adaptive $z_{\kf}$ factor [9].}
\label{pbf}
\end{figure} 

Although superfluidity strongly suppresses the standard neutrino emission
channels, there is a powerful mechanism for neutrino emission in 
superfluid matter due to the PBF process~\cite{Flowers}. This is a 
significant source for neutrinos from the density range, where the
temperature lies between $0.2 \, T_c(\kfn) \lesssim T < T_c(\kfn)$.
The emissivity from the $^1$S$_0$ neutron pair-breaking and formation
process is given by
\beq
\varepsilon_{\text{PBF}} = 1.170 \cdot 10^{21} \, N_\nu \, \frac{m_n^\ast}{m_n}
\, \frac{\kfn}{m_n} \, T_9^7 \, F(\tau) \,\text{erg cm}^{-3} \, 
\text{s}^{-1} ,
\eeq
where the function $F(\tau)$ depends on the critical temperature
$\tau = T/T_c(\kfn)$ at given neutron Fermi momentum (for S-wave 
pairing $T_c(\kfn) = 0.57 \Delta(\kfn)$). Details and a
parametrization of $F(\tau)$ which we employ can be found 
in~\cite{Yakovlev}. In Fig.~\ref{pbf} we compare our results for the
bremsstrahlung emissivity in the normal state to the $^1$S$_0$ PBF
process. While it demonstrates that the PBF process is extremely effective
at higher temperatures $T \gtrsim 10^{8.5} \, \text{K}$, at lower
temperatures the density range for emission via PBF is rather narrow and
we expect the volume-integrated bremsstrahlung emission to dominate.
Fig.~\ref{pbf} also nicely illustrates how the cooling of neutron stars is
able to probe the internal structure, with the enhanced emissivity
from the PBF process serving as a clear signal of the superfluid
phase.

We finally note that at higher neutron densities the proton PBF process
is also in effect. However, at lower temperatures the cooling through neutrino
bremsstrahlung in normal-state $nn$ collisions will be larger for the same
reason as above, and in addition due to the smaller number of protons.

\section{Summary}
\label{conclude}

We have computed the emissivity from neutrino pair bremsstrahlung in 
nucleon-nucleon collisions in pure neutron matter, within an effective 
theory of quasiparticle interactions in the vicinity of the Fermi 
surface. The effective scattering amplitude is calculated from the 
model-independent low-momentum nucleon-nucleon interaction $\vlk$
to second order, keeping the full non-central spin dependence~\cite{AB}.
We find that inclusion of medium modifications, in particular the
renormalization of the tensor force, reduces the emissivity 
compared to the tree-level $\vlk$ result by a multiplicative
factor $0.64$ at nuclear 
matter density (or a factor $0.5$ relative to the direct one-pion
exchange estimate). At sub-nuclear densities, the reduction is $0.2$ at 
a fifth of nuclear matter density.  Furthermore, we find that the 
effect of spin non-conserving parts in the scattering amplitude is 
rather small. While the temperature dependence of the emissivity is 
naturally very important for cooling simulations, the density 
dependence needs to be considered as well, since the luminosity of 
the neutron star involves an integral over the volume of the star.

When polarization effects are included in the $^3$P$_2$ neutron pairing 
interaction, a considerable reduction of the superfluid gaps was 
found~\cite{AB}. Therefore, the neutrino bremsstrahlung process could 
be more important for the cooling of neutron stars than believed, since 
superfluidity of protons suppresses Urca processes as well as 
bremsstrahlung in $np$ and $pp$ scattering, whereas neutrons remain
in the non-superfluid phase at higher densities at typical 
core temperatures in the neutrino cooling stage. Even at lower densities,
where neutrons form a $^1$S$_0$ superfluid, the lore is that neutrino 
bremsstrahlung from neutrons dominates the modified Urca process in 
the presence of superfluidity~\cite{Yakovlev}, although the PBF
channel is considerably more effective. We also note that the 
bremsstrahlung process, in contrast to the Urca channels, produces
$\nu_\mu$  and $\nu_\tau$ neutrinos. In order to compare our results for
the bremsstrahlung rates to the emission through the PBF process, we have
shown results for the PBF process for realistic pairing gaps obtained in
the RG approach~\cite{RGnm}. This nicely demonstrates that the PBF channel
is very effective and can act as a powerful signal of superfluidity for 
temperatures comparable to the maximal critical temperature, but is
restricted only to a narrow density range for lower temperatures.
In the latter regime, one thus expects the integrated emissivity from 
bremsstrahlung to dominate.

Our results provide neutrino bremsstrahlung rates derived from successfully
used effective nuclear interactions. The emissivities can be used as 
microscopic input for neutron star cooling simulations, in conjunction with the
superfluid gaps calculated in~\cite{AB,RGnm}. A RG calculation of effective
interactions in asymmetric matter, which will address both the effects of
induced interactions on proton pairing in neutron star matter as well as
higher-order contributions to non-central interactions, is in 
preparation~\cite{prep}. A self-consistent treatment of the tensor force
is necessitated by a substantial renormalization at second order. Such
studies will further constrain neutrino emissivities microscopically in a 
consistent framework. Our results can then be incorporated in cooling 
simulations of neutron stars to possibly constrain the structure of 
neutron stars and their densest interiors.

\section*{Acknowledgments} 

We thank Bengt Friman and Dick Furnstahl for useful discussions. 
AS is supported by the NSF under Grant No. PHY-0098645 and through an 
Ohio State University Postdoctoral Fellowship. The work of PJ and 
CG is supported jointly by the Natural Sciences and Engineering 
Research Council of Canada and the Fonds Nature et Technologies of Quebec.

\end{document}